\documentclass[aps, prl,twocolumn, noshowpacs, groupedaddress,nopacs,floatfix]{revtex4}

\usepackage{graphicx}
\usepackage{color}
\usepackage{amssymb}
\usepackage{amsmath}
\usepackage{times}
\usepackage[normalem]{ulem}

\begin{document}
\title{Superradiance for atoms trapped along a photonic crystal waveguide}

\author{A. Goban$^{1,2,\dag}$, C.-L. Hung$^{1,2,\dag,\ddag}$, J. D. Hood$^{1,2,\dag}$, S.-P. Yu$^{1,2,\dag}$\footnotetext{\small $^{\dag}$These authors contributed equally to this research.},\\
J. A. Muniz$^{1,2}$, O. Painter$^{2,3}$, and H. J. Kimble$^{1,2,\ast}$\footnotetext{\small $^{\ddag}$ Present address: Purdue University, West Lafayette, IN 47906, USA} \footnotetext{\small $^{\ast}$ Correspondence and requests for materials should be addressed to HJK (hjkimble@caltech.edu.)}}

\address{$^1$ Norman Bridge Laboratory of Physics 12-33}
\address{$^2$ Institute for Quantum Information and Matter, California Institute of Technology, Pasadena, CA 91125, USA}
\address{$^3$ Thomas J. Watson, Sr., Laboratory of Applied Physics 128-95}

\begin{abstract}
We report observations of superradiance for atoms trapped in the near field of a photonic crystal waveguide (PCW). By fabricating the PCW with a band edge near the D$_1$ transition of atomic cesium, strong interaction is achieved between trapped atoms and guided-mode photons. Following short-pulse excitation, we record the decay of guided-mode emission and find a superradiant emission rate scaling as $\bar{\Gamma}_{\rm SR}\propto\bar{N}\cdot\Gamma_{\rm 1D}$ for average atom number $0.19 \lesssim \bar{N} \lesssim 2.6$ atoms, where $\Gamma_{\rm 1D}/\Gamma_0 =1.1\pm0.1$ is the peak single-atom radiative decay rate into the PCW guided mode and $\Gamma_{0}$ is the Einstein-$A$ coefficient for free space. These advances provide new tools for investigations of photon-mediated atom-atom interactions in the many-body regime.
\end{abstract}

\date{\today}
\maketitle


Interfacing light with atoms localized near nanophotonic structures has attracted increasing attention in recent years. Exemplary experimental platforms include nanofibers \cite{Vetsch10, Goban12}, photonic crystal cavities \cite{Thompson13} and waveguides \cite{Yu14,Goban14}. Owing to their small optical loss and tight field confinement, these nanoscale dielectric devices are capable of mediating long-range atom-atom interactions using photons propagating in their guided modes. This new paradigm for strong interaction of atoms and optical photons offers new tools for scalable quantum networks \cite{Kimble08}, quantum phases of light and matter \cite{Plenio06,Greentree06}, and quantum metrology \cite{Komar14}.

In particular, powerful capabilities for dispersion and modal engineering in nanoscopic photonic crystal waveguides (PCWs) provide opportunities beyond conventional settings in AMO physics within the new field of \textit{waveguide QED} \cite{Chang2007b, Vetsch10, Goban12,Goban14, Wallraff13,Mlynek14}. For example, the edge of a photonic band gap aligned near an atomic transition strongly enhances single-atom emission into the one-dimensional (1D) PCW due to a van-Hove singularity at the band edge (i.e., a `slow-light' effect \cite{baba08, Hung13,Lodahl14}). Because the Bloch function for a guided mode near the band edge approaches a standing-wave, symmetric optical excitations can be induced in an array of trapped atoms, resulting in superradiant emission \cite{Dicke54, Gross1982} into the PCW. Superradiance has important applications for realizing quantum memories \cite{Duan01, Kuzmich03, vanderWal03, Casabone15,Reimann15}, single photon sources \cite{Chou04, Black05}, laser cooling by way of cooperative emission \cite{Chan03, Wolke12}, and narrow linewidth lasers \cite{Bohnet12}. Related cooperative effects are predicted in nano-photonic waveguides absent an external cavity~\cite{Kien2005a}, including atomic Bragg mirrors \cite{Chang12} and self-organizing crystals of atoms and light \cite{Deutsch1995, chang2013, grieser2013}. 

Complimentary to superradiant emission is the collective Lamb shift induced by proximal atoms virtually exchanging off-resonant photons \cite{Svidzinsky08, Scully09Lamb, Rohlsberger10, Keaveney12}. With the atomic transition frequency placed in a photonic band gap of a PCW, real photon emission is largely suppressed. 
Coherent atom-atom interactions then emerge as a dominant effect for QED with atoms in bandgap materials~\cite{Kurizki90, John90, Shahmoon13, Lambropoulos2000,Douglas13,Tudela14}. Both the strength and length scale of the interaction can be `engineered' by suitable band shaping of the PCW, as well as dynamically controlled by external lasers \cite{Douglas13, Tudela14}. Exploration of many-body physics with tunable and strong long-range atom-atom interactions are thereby enabled~\cite{Douglas13,Tudela14}.

In this Letter, we present an important advance for the field of waveguide QED. We describe an experiment that cools, stably traps, and interfaces multiple cold atoms along a quasi one-dimensional PCW. Through precise band edge alignment and guided-mode (GM) design, we achieve strong radiative coupling of one trapped atom and a GM of the PCW, such that the inferred single-atom emission rate into the GM is $\Gamma_{\rm 1D}/\Gamma_0 = 1.1 \pm0.1$, where $\Gamma_{\rm 1D}$ is the peak single-atom radiative decay rate into the PCW guided mode and $\Gamma_0$ is the Einstein-$A$ coefficient for free space. With multiple atoms, we observe superradiant emission in both time and frequency domains with measurements of transient decay following pulsed excitation and steady-state transmission spectra, respectively. We infer cooperative, superradiant coupling with rate $\bar{\Gamma}_{\rm SR}$ that scales with the mean atom number $\bar{N}$ as $\bar{\Gamma}_{\rm SR} = \eta\bar{N}\cdot\Gamma_{\rm 1D}$ over the range $0.19\lesssim \bar{N} \lesssim 2.6$ atoms, where $\eta=0.34\pm0.06$.

\begin{figure}[b!]
\centering
\vspace{-4mm}
\includegraphics[width=1.0\columnwidth]{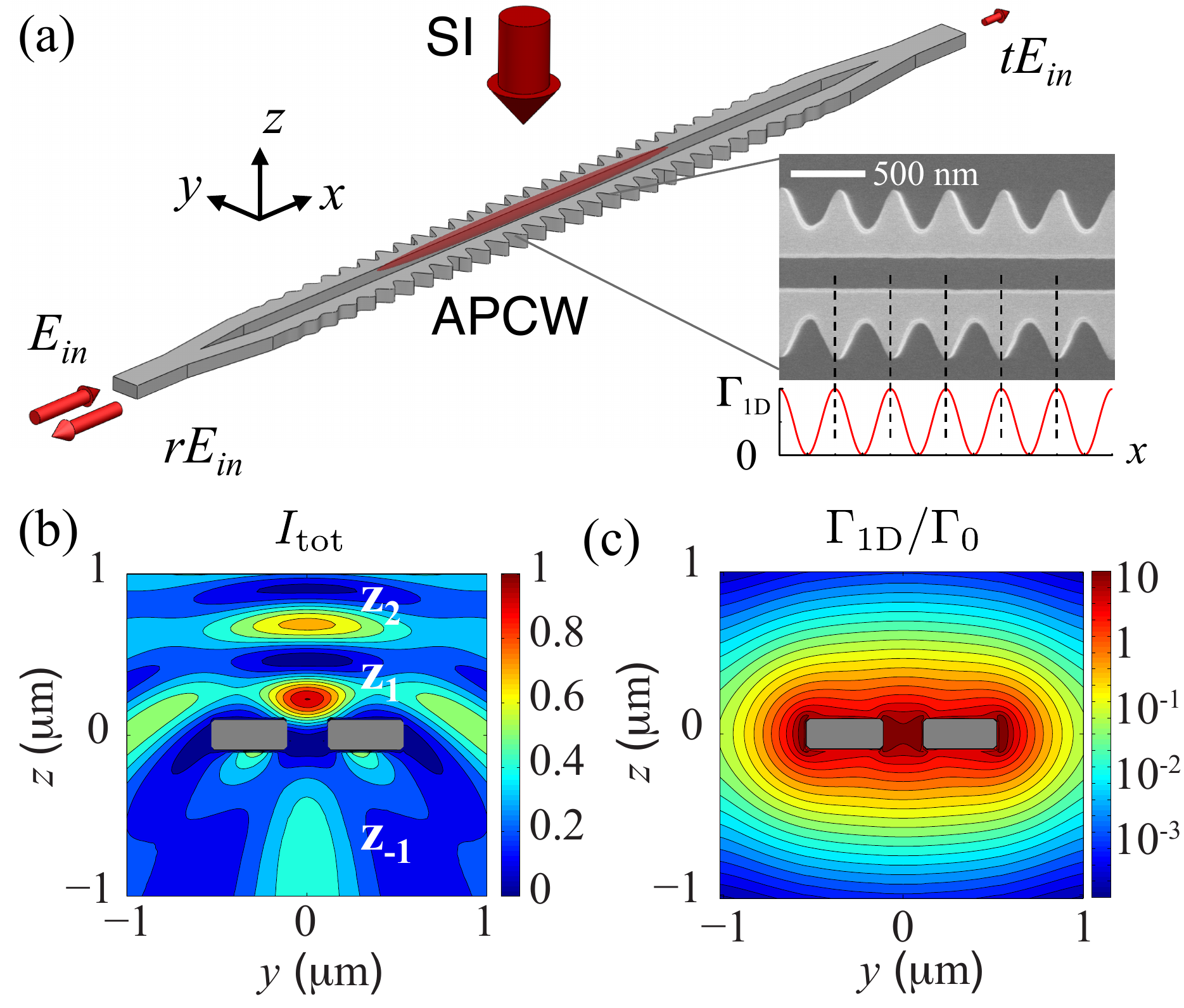}
\vspace{-5mm}
\caption{Trapping and interfacing atoms with a 1D photonic crystal waveguide. (a) A side-illumination (SI) beam is reflected from an `alligator' photonic crystal waveguide (APCW) to form a dipole trap to localize atoms near the APCW (gray shaded structure). The red shaded region represents trapped atoms along the APCW. An incident field $E_{\rm in}$  excites the TE-like fundamental mode and thereby trapped atoms couple to this guided mode (GM). The transmitted $tE_{\rm in}$ and reflected field $rE_{\rm in}$ are recorded.
The inset shows an SEM image of the APCW and corresponding single-atom coupling rate $\Gamma_{\rm 1D}$ along the $x$ axis at the center of the gap ($y=0$). (b) Normalized intensity cross section of the total intensity $I_{\rm tot}$ resulting from the SI beam and its reflection, which form an optical dipole trap. Trap locations along the $z$ axis at $y=0$ are marked by $z_i$. Masked gray areas represent the APCW. (c) The single-atom coupling rate into the TE guided mode $\Gamma_{\rm 1D}(0,y,z)$ normalized to the free-space decay rate $\Gamma_0$ for the cesium D$_1$ line.}
\label{fig1}%
\end{figure}

Our experimental platform is based on trapped cesium atoms near a 1D alligator photonic crystal waveguide (APCW) \cite{Yu14,Goban14}. The APCW is formed by two parallel SiN nanobeams separated by 238 nm with periodic corrugations at the outer edges (Fig.~1(a)). The APCW consists of $150$ identical unit cells with lattice constant $a=371$nm (length $L \simeq 55.7\mu$m) and is terminated at either end by $30$ tapered cells for mode matching to parallel nanobeams without corrugation. 
Photons can be coupled into and out of the APCW from conventional cleaved-fibers at either end of the structure. Design principles, fabrication methods, and device characterization of the APCW can be found in Refs. \cite{Yu14, Goban14,SM}.

For the APCW used here, we align the band edge of the fundamental guided mode (electric field predominantly transverse-electric (TE) polarized in the plane of the waveguide) near the cesium D$_1$ line at 894.6 nm, with a mode-matched TE input field $E_{\rm in}$ tuned around the $6S_{1/2},~F=3 \rightarrow 6P_{1/2},~F'=4$ transition.
Near the band edge, the atom-photon coupling rate is significantly enhanced by the group index $n_g$, as well as by reflections from the tapering regions that surround the APCW. From the measured transmission spectrum of the device absent atoms, we estimate a group index $n_g \simeq 11$ and an intensity enhancement $\mathcal{E}_I\sim6$ from the taper reflections~\cite{SM}.

To trap atoms along the APCW, we create tight optical potentials using the interference pattern of a side-illumination (SI) beam and its reflection from the surface of the APCW \cite{Thompson13, SM}.
The polarization of the SI beam is aligned parallel to the $x$-axis of the 1D waveguide to maximize the reflected field. Figure~\ref{fig1}(b) shows the calculated near-field intensity distribution in the $y$-$z$ plane \cite{COMSOL}.
With a red-detuned SI beam, cold atoms can be localized to intensity maxima (e.g., positions $z_{-1},z_{1},z_{2}$ in Fig.~\ref{fig1}(b)).
However, because of the exponential falloff of the GM intensity, only those atoms sufficiently close to the APCW can interact strongly with guided-mode photons of the input field $E_{\rm in}$, Fig. \ref{fig1}(c).
The trap site with the strongest atom-photon coupling is located at $( y_1, z_1 ) =(0,220)$~nm, closest to the center of the unit cell and $\Delta z\sim120$ nm from the plane of the upper surfaces of the APCW.
Other locations are calculated to have coupling to the fundamental TE-like mode less than $1\%$ of that for site $z_1$ (e.g., the sites at $z_{-1},z_{2}$ have intensity ratios $I(z_{-1})/I(z_{1})=0.01, I(z_{2})/I(z_{1})=0.005$). 

Along the $x$ axis of the APCW, the dipole trap $U(x,0,z_1)$ is insensitive to the dielectric corrugation within a unit cell and is nearly uniform to within $<2\%$ around the central region of the APCW. By contrast, atom emission into the fundamental TE-like mode is strongly modulated with $\Gamma_{\rm 1D} (x,0,z_1)\simeq\Gamma_{\rm 1D}\cos^2(kx)$ due to the Bloch mode function near the band edge of the APCW ($k\approx\pi/a$), as shown in the inset of Fig.~\ref{fig1}(a). 
Thus, even for atoms uniformly distributed along the $x$ axis of the trapping potential, only those close to the center of a unit cell can strongly couple to the guided mode, greatly facilitating phase-matched symmetric excitation of the atoms. In our experiment, we have chosen a 50~$\mu$m waist for the SI beam to provide weak confinement along the $x$ axis, with atoms localized near the central region ($\Delta x \simeq \pm 10~\mu$m) of the APCW for the estimated temperature $\sim 50 \mu$K from a time-of-flight measurement in free space. The SI beam for dipole trapping is 220~GHz red-detuned with respect to the D$_2$ line and has a total power of 50~mW for all measurements reported. 

\begin{figure}[t]
\centering
\includegraphics[width=1\columnwidth]{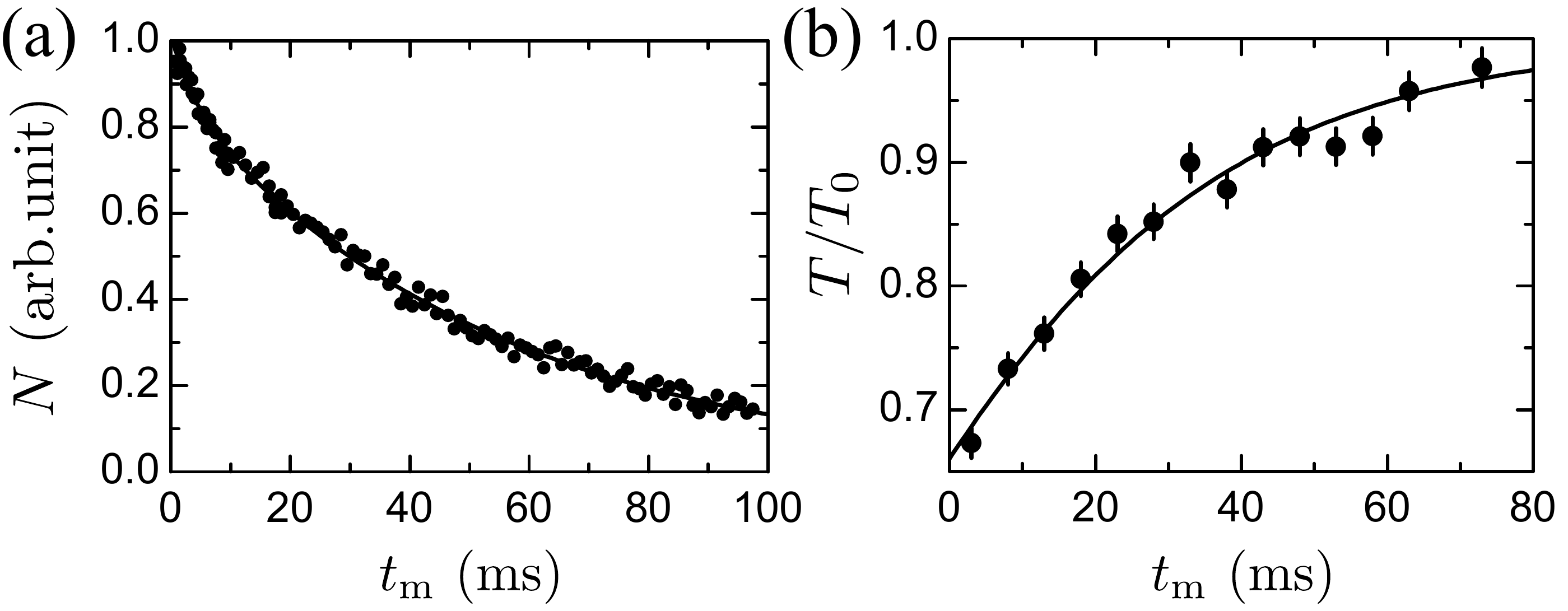}
\vspace{-5mm}
\caption{Lifetime of trapped atoms near the APCW. (a)  $1/e$-lifetime of $\tau_{\rm fs}= 54\pm5~$ms is determined using free-space absorption imaging of the trapped atom cloud. (b) $1/e$-lifetime of $\tau_{\rm GM}=28\pm2$~ms is observed from the normalized transmission $T/T_0$ of resonant GM probe pulses.}
\vspace{-5mm}
\label{fig2}%
\end{figure}

Cold atoms from a MOT that surrounds the APCW \cite{Goban14} are loaded into the dipole trap during an optical molasses phase ($\sim 5$ ms) and then optically pumped to $6S_{1/2}$, $F=3$ for $\sim$1 ms. Atoms are held in the dipole trap for time $t_{\rm hold}$ relative to the end of the loading sequence, and then free-space absorption imaging is initiated over the interval ($t_{\rm hold},~t_{\rm hold}+\Delta t_{\rm m}$) with $\Delta t_{\rm m}=0.2$ ms. We introduce the measured time $t_{\rm m}=t_{\rm hold}+\Delta t_{\rm m}/2$, centered in the measurement window. As shown in Fig. \ref{fig2}(a), we measure a trap lifetime $\tau_{\rm fs}=54\pm5$ ms and find a peak density $\rho_0 \approx 2\times 10^{11}~$cm$^{-3}$ near the APCW. The atom density $\rho$ near the APCW can be adjusted over a wide range $0.06 \lesssim\rho/\rho_0 \leqslant 1$ by varying the duration of the MOT loading cycle while keeping all other procedures identical. 

To determine the lifetime for trapped atoms near the APCW, we again hold atoms for $t_{\rm hold}$, and then launch $E_{\rm in}$ as a resonant GM probe in measurement interval $t_{\rm m}\pm\Delta t_{\rm m}/2$ with $\Delta t_{\rm m} =5$ ms. From the recorded transmitted signals, we compute $T/T_0$, where $T_0$ is the transmission without atoms. During the probe period, we also apply free-space repump beams, tuned to the D$_2$, $6S_{1/2},~F=4 \rightarrow 6P_{3/2},~F'=4$ resonance, to remove population in the $6S_{1/2}$, $F=4$, since the probe excites an open transition.  Fig. \ref{fig2}(b) shows $T/T_0$ gradually recovering to $T/T_0=1$ as $t_{\rm m}$ increases, with a fit to the data giving a $1/e-$time of $\tau_{\rm GM}=28\pm2~$ms~\cite{SM}. $\tau_{\rm GM}$ is consistently shorter than $\tau_{\rm fs}$ from free-space imaging, which might be attributed to increased heating from the stronger light intensity near the APCW, the effect of surface potentials, or outgassing from the silicon chip and structures that support the APCW. These contributions are being investigated in more detail.

Our principal investigation of superradiance involves observation of the transient decay of emission from an array of atoms trapped along the APCW. For a collection of $N>1$ atoms, superradiance is heralded by a total decay rate $\Gamma_{\rm tot}=\Gamma_{\rm SR}+\Gamma^{(1)}_{\rm tot}$ that is enhanced beyond the total rate of decay for one atom $\Gamma^{(1)}_{\rm tot}=\Gamma_{\rm 1D}+\Gamma'$. $\Gamma_{\rm SR}$ is the $N$-dependent superradiant rate operationally determined from $\Gamma_{\rm tot}$ and $\Gamma^{(1)}_{\rm tot}$. Here, $\Gamma'$ is the radiative decay rate into all channels other than the TE-like GM of the APCW.
We numerically evaluate $\Gamma'/\Gamma_0\approx1.1$ for an atom at the trap site $z_1$ in Fig. \ref{fig1}(b) along the APCW, with $\Gamma_0$ the free-space decay rate for the D$_1$ transition~\cite{Hung13, Rafac1999}.
Cooperative level shifts $|H_{\rm dd}| \ll \Gamma_{\rm 1D}$ are neglected for the current configuration of our experiment \cite{SM}.

\begin{figure}[t!]
\centering
\includegraphics[width=.85\columnwidth]{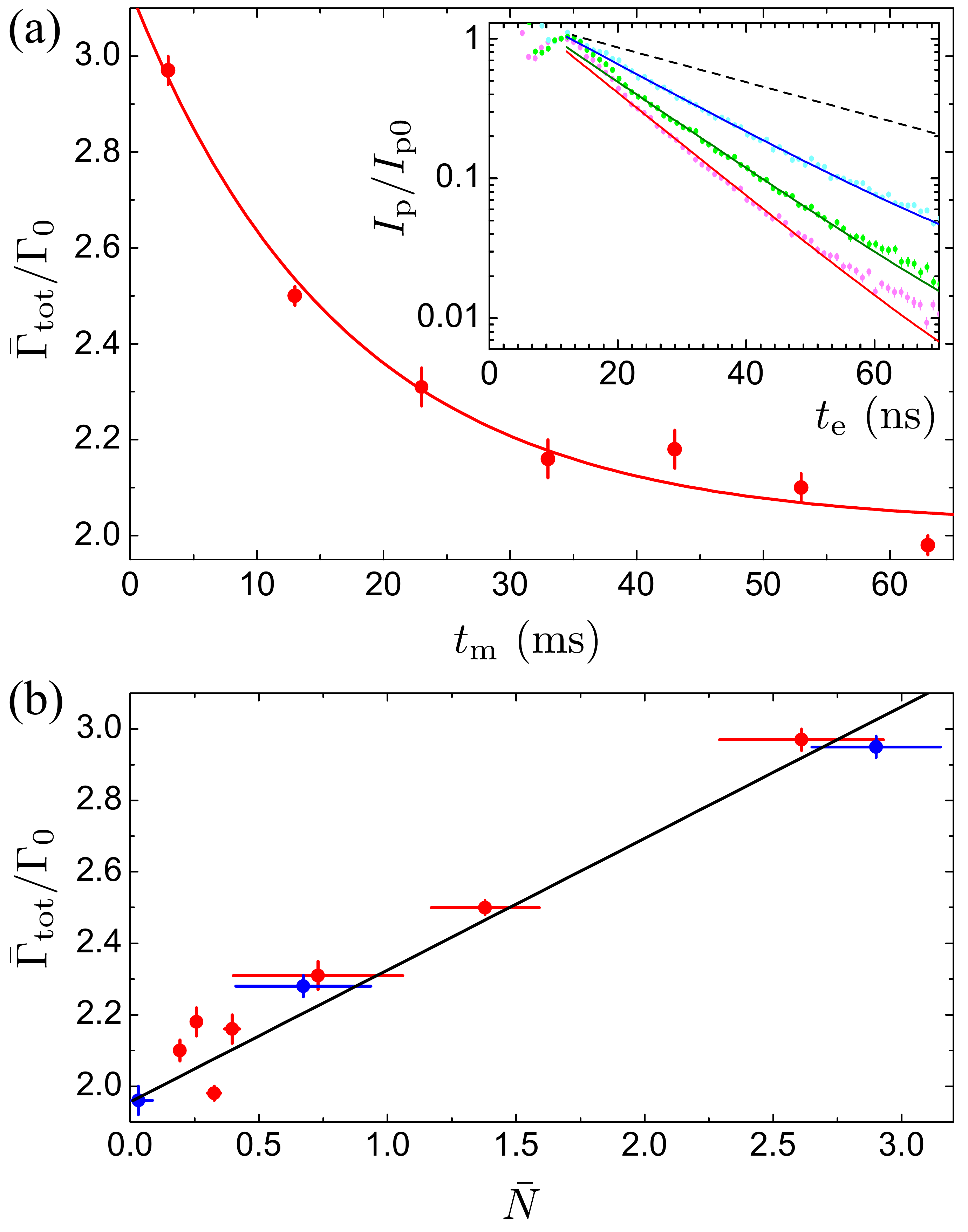}
\vspace{-2mm}
\caption{Decay rate and atom number dependence (a) Fitted total decay rate $\bar{\Gamma}_{\rm tot}$ normalized with free-space decay rate $\Gamma_0$ (circles) as a function of measurement time $t_{\rm m}$. The solid line is a simple exponential fit to determine the superradiant decay rate $\bar{\Gamma}_{\rm SR}/\Gamma_0=1.1\pm0.1$ and the single-atom decay rate $\bar{\Gamma}_{\rm tot}^{(1)}/\Gamma_0=2.0\pm0.1$ with $\tau_{\rm SR}=17\pm3$ ms. The inset shows the temporal profiles of normalized guided-mode emission $I_{\rm p}/I_{\rm p0}$ (circles) with  $I_{\rm p0}$ the peak emission. Exponential fits (solid curves): $t_{\rm m}=3$ ms (red), 13 ms (green),  and 63 ms (blue).
The black dashed curve shows a exponential decay with free-space decay rate $\Gamma_0$.
(b) Fitted total decay rate $\bar{\Gamma}_{\rm tot}$ normalized with $\Gamma_0$ as a function of mean number of trapped atoms $\bar{N}$ from a detailed model \cite{SM}.
We adjust $\bar{N}$ by changing the trap hold time (red circles) or atom loading time (blue circles). The black line is a linear fit to the combined data sets, giving $\bar{\Gamma}_{\rm SR}=\eta\cdot\bar{N}\cdot\Gamma_{\rm 1D}$ with $\eta=0.34\pm0.06$.}
\vspace{-5mm}
\label{fig3}%
\end{figure}

We record the temporal profiles of atomic emission into the fundamental TE-like GM following short-pulse ($\sim$10~ns FWHM), resonant excitations via $E_{\rm in}$. To ensure small population in the excited state, we choose a pulse intensity well below the saturation intensity ($I/I_{\rm sat}<0.1$).
After a time $t_{\rm hold}$ the excitation cycle is repeated every $500$ ns for $\Delta t_{\rm m}=6$ ms, and detection events are accumulated for the reflected intensity $|rE_{\rm in}|^2$ by an avalanche photodiode (APD).
We consider decay curves of GM emission at $15~{\rm ns}<t_{\rm e}<70$ ns after the center of the excitation pulse
(i.e., after the excitation pulse is sufficiently extinguished, $t_{\rm e} >15$ ns, and while the background counts are negligible compared to the atomic emission, $t_{\rm e} \lesssim 70$ ns \cite{SM}). The total decay rate $\bar{\Gamma}_{\rm tot}$ is extracted by simple exponential fits as shown in the inset of Fig. \ref{fig3}(a). The deviation from the exponential fit at $t_{\rm e}\gtrsim60$ ns is due to the spatially varying coupling rate $\Gamma_{\rm 1D}\cos^2(kx)$, which is captured by a detailed model discussed later~\cite{SM}.

Enhanced total decay rate with increasing atom number is clearly evidenced in Fig.~\ref{fig3}(a), where the atom number can be adjusted by varying trap hold time $t_{\rm hold}$ prior to the measurement.
At the shortest measurement time $t_{\rm m}=3$ ms with $t_{\rm hold} =0$ ms (i.e., the maximum number of trapped atoms), the measured total decay rate is largest at $\bar{\Gamma}_{\rm tot}/\Gamma_0\approx2.9$. At $t_{\rm m}=63$ ms much longer than the trap lifetime $\tau_{\rm GM}=28\pm2~$ms, the total decay rate settles to $\bar{\Gamma}_{\rm tot}/\Gamma_0\approx2.0$. This asymptotic behavior suggests that $\bar{\Gamma}_{\rm tot}$ at long hold time corresponds to the single-atom decay rate $\bar{\Gamma}^{(1)}_{\rm tot}$.

To determine quantitatively the superradiant and single-atom emission rates from our measurements of decaying GM emission, we present two different analyses that yield consistent results. First is a simple and intuitive analysis applied to Fig. 3(a) in which we employ an empirical exponential fit, $\bar{\Gamma}_{\rm tot}(t_{\rm m})=\bar{\Gamma}_{\rm SR}e^{-t_{\rm m}/\tau_{\rm SR}} + \bar{\Gamma}^{(1)}_{\rm tot}$, with the superradiant $\bar{\Gamma}_{\rm SR}$, single-atom $\bar{\Gamma}^{(1)}_{\rm tot}$, and $\tau_{\rm SR}$ characterizing decay of superradiance due to the atom loss. The fit yields the maximum superradiant rate $\bar{\Gamma}_{\rm SR}/\Gamma_0=1.1\pm0.1$ with $\tau_{\rm SR}=17\pm3$ ms, and a reasonable correspondence to the measured decay rates $\bar{\Gamma}_{\rm tot}$, as shown by the red curve in Fig.~\ref{fig3}(a). The asymptote $\bar{\Gamma}^{(1)}_{\rm tot}/\Gamma_0 = 2.0\pm0.1$ gives the total single-atom decay rate. With $\Gamma'/\Gamma_0\approx1.1$ determined numerically for an atom at trap site $z_1$ along the APCW (Fig. \ref{fig1}(b)), we deduce $\bar{\Gamma}_{\rm 1D}/\Gamma_0 =0.9\pm0.1$ for the single-atom decay rate into the GM of the APCW. 

To substantiate this simple emphirical model, our second analysis is a detailed number treatment based upon transfer matrix calculations \cite{SM}.
Decay curves are generated for a fixed number of atoms $N$ distributed randomly along along the x-axis of the APCW with uniform probability density but with spatially varying spatially varying coupling $\Gamma_{\rm 1D}(x)\simeq\Gamma_{\rm 1D} \cos^2(kx)$.
These $N$-dependent, spatially-averaged decay curves are further averaged over a Poisson distribution with mean atom number $\bar{N}$, capturing the variation of atom number $N$ as we repeat experiments for data accumulation. Fitting to this model, we extract $\Gamma_{\rm 1D}/\Gamma_0=1.1\pm0.1$ for measurements at long hold time (e.g., at $t_{\rm m}=63$ms in Fig. \ref{fig3}(a)).
Since the intensity of the fluorescence from a single atom is spatially modulated by $\cos^4(kx)$, only an atom near the center of unit cell can strongly couple to the GM, resulting in the small difference between averaged $\bar{\Gamma}_{\rm 1D}$ and peak $\Gamma_{\rm 1D}$.
Also, the decay curve for GM emission at $t_{\rm m}=3$ ms can be well fitted with $\bar{N}=2.6\pm0.3$ atoms~\cite{SM}.
The red points in Fig. \ref{fig3} (b) display the total decay rate $\bar{\Gamma}_{\rm tot}$ as a function of $\bar{N}$ extracted from fits of the transfer matrix model to the measured decay curves, which clearly shows that superradiance emission rate is proportional to $\bar{N}$.

The value $\Gamma_{\rm 1D}/\Gamma_0=1.1\pm0.1$ from our measurements agrees reasonably well with the theoretical value $\Gamma_{\rm 1D}/\Gamma_0 \approx1.2$ determined by FDTD calculations \cite{meep, SM}, despite several uncertainties (e.g., locations of trap minima relative to the APCW with uncertainty below $10$ nm). The  agreement validates the absolute control of our fabrication process (including the negligible effect of loss and disorder along the APCW), as well as the power of the theoretical tools that we have developed \cite{Hung13,Douglas13,Tudela14}.

We confirm that the variation of $\bar{\Gamma}_{\rm tot}$ in Fig. \ref{fig3}(a) is not due to the heating of atomic motion during the trap hold time. To see this, we adjust $\bar{N}$ via different MOT loading times and measure the decay rate at the shortest hold time ($t_{\rm m}=3$ ms), as shown by blue points in Fig. \ref{fig3}(b) . These observations are consistent with those from varying the trap hold time (red points in Fig. \ref{fig3}(b)), and lead to an almost identical single-atom decay rate $\bar{\Gamma}^{(1)}_{\rm tot}/\Gamma_0=2.0\pm0.1$ at the shortest loading time, corresponding to $\rho/\rho_0 =0.16$ and $\bar{N} \ll 1$. 

The data and our analysis related to Fig. \ref{fig3} strongly support the observation of superradiant decay for atoms trapped along the APCW. Assuming $\bar{\Gamma}_{\rm tot} = \bar{\Gamma}_{\rm SR}+ \bar{\Gamma}_{\rm tot}^{(1)}$ and fitting $\bar{\Gamma}_{\rm tot}$ linearly with $\bar{N}$, as shown in Fig. \ref{fig3}(b), we find that the superradiant rate is given by $\bar{\Gamma}_{\rm SR} = \eta\cdot\bar{N}\cdot\Gamma_{\rm 1D}$ with $\eta=0.34\pm0.06$.
The slope $\eta$ is reduced below unity by the random distribution of atoms along the $x$-axis \cite{SM}.

\begin{figure}[t!]
\centering
\includegraphics[width=.85\columnwidth]{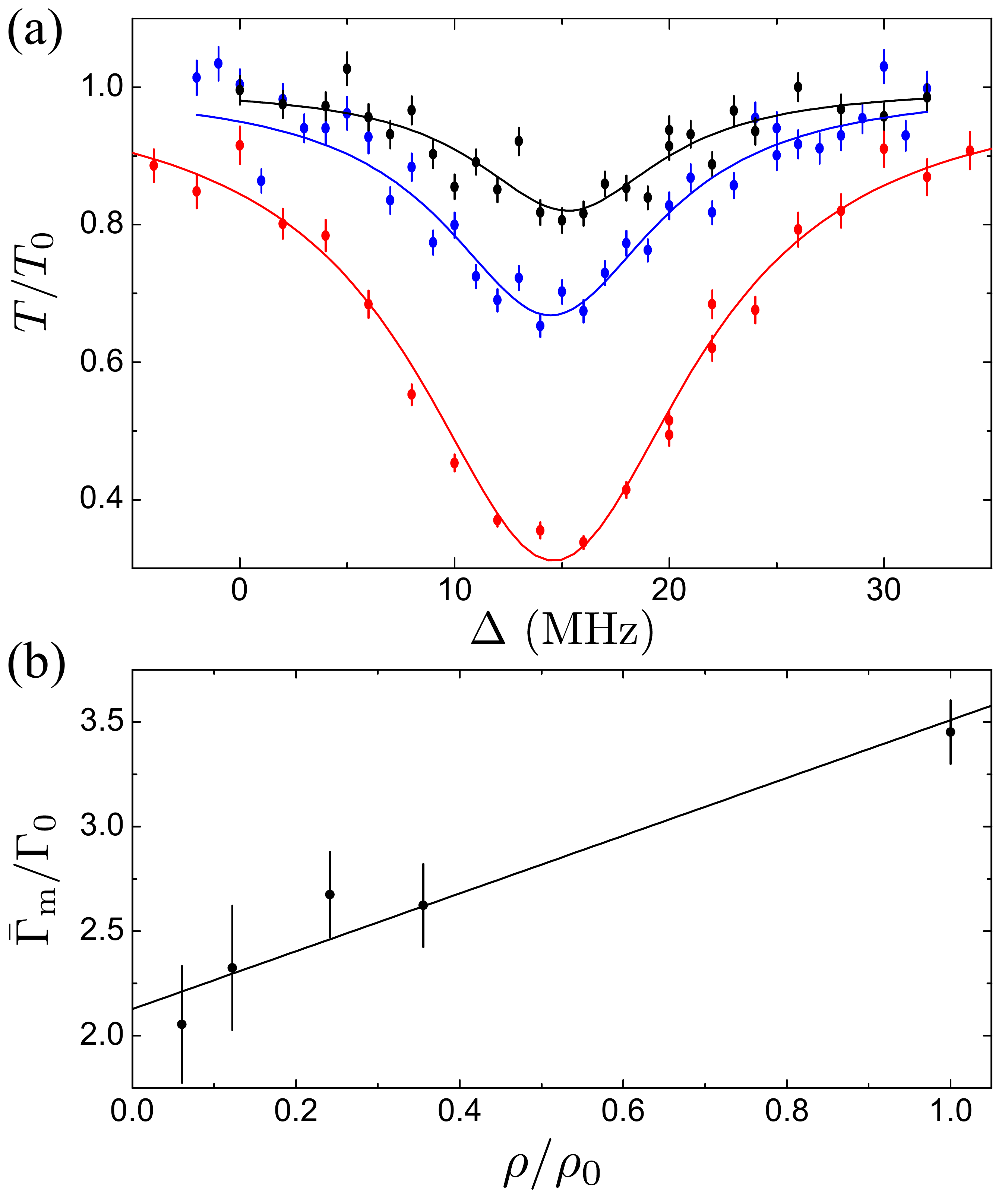}
\vspace{-2mm}
\caption{Steady-state transmission spectra $T(\Delta)$ and fitted atomic linewidth $\bar{\Gamma}_{\rm m}$. (a) $T(\Delta)$ with $\Delta=0$ corresponding to the free-space line center. The three sets of points are measured at relative densities $\rho/\rho_0= 0.12$ (black), 0.24 (blue), and 1 (red), where the transmission without atoms is $T_0$. Solid curves are Lorentzian fits to determine the linewidth $\bar{\Gamma}_{\rm m}$. Each point in the spectra is an average over 10 experiment repetitions.  (b) Fitted linewidths (circles) normalized to  $\Gamma_0$ as a function of $\rho/\rho_0$. The solid line is a linear fit with intercept of $\bar{\Gamma}^{(1)}_{\rm m}/\Gamma_0=2.1\pm0.1$.}
\vspace{-5mm}
\label{fig4}
\end{figure}

This observation of superradiant decay is complemented by line broadening for steady-state transmission spectra $T(\Delta)$ measured at $t_{\rm m}=3$ ms with $\Delta t_{\rm m}=5$ ms, as show in Fig. \ref{fig4}.
The measured linewidths $\bar{\Gamma}_{\rm m}$ are significantly broader than the free-space width (FWHM) $\Gamma_0/2\pi = 4.56$ MHz~\cite{Rafac1999}, predominantly due to cooperative atomic coupling to the GM of the APCW.
We also observe a significant drop in $T/T_0$ at line center due to strong atom-photon coupling. 
Indeed, in Fig.~\ref{fig4} (a), we measure $T/T_0 \simeq 0.30$ (i.e., a $70\%$ attenuation of the GM flux $|E_{\rm in}|^2$) for maximum density $\rho_0$, and $T/T_0 \simeq 0.95$ at the lowest density investigated, $\rho/\rho_0\approx 0.06$. 

No clear density dependent shift is observed in Fig. \ref{fig4}(a), in support of our neglect of cooperative energy shifts $|H_{\rm dd}|$ \cite{SM}. 
The shift in line center for $T(\Delta)$ from $\Delta = 0$ in free space to $\Delta = 14~$MHz for atoms trapped along the APCW is induced by the dipole trap.
Furthermore, trapped atoms should suffer small inhomogeneous broadening in the spectra shown in Fig.~\ref{fig4}, since the FORT shift is small ($<$ 1 MHz) for the $6P_{1/2}, F=4'$ excited state, and atoms are well localized around the trap center due to their low temperature $T \sim 50 \mu$K, corresponding to a small range of light shifts $\lesssim 1$MHz for atoms in the ground state.

In Fig.~\ref{fig4} (b), we plot the linewidths $\bar{\Gamma}_{\rm m}$ extracted from $T(\Delta)$ as a function of $\rho/\rho_0$. $\bar{\Gamma}_{\rm m}/\Gamma_0\approx 3.4$ is largest at $\rho/\rho_0=1$, and reduces to $\bar{\Gamma}_{\rm m}/\Gamma_0\approx 2.1$ at $\rho/\rho_0=0.06$. From linear extrapolation, the single-atom linewidth is estimated to be $\bar{\Gamma}^{(1)}_{\rm m}/\Gamma_0=2.1\pm0.1$. Absent inhomogeneous broadening, we expect that $\bar{\Gamma}^{(1)}_{\rm m} = \bar{\Gamma}_{\rm 1D} + \Gamma'$. With the calculated $\Gamma'/\Gamma_0\approx1.1$, the single-atom coupling rate can be simply deduced as $\bar{\Gamma}_{\rm 1D}/\Gamma_0 \approx 1.0 \pm 0.1$. 
A simple estimate of the maximum mean number of atoms then follows from $\bar{N}_{\rm m} = (\bar{\Gamma}_{\rm m}(\rho_0) - \Gamma')/\bar{\Gamma}_{\rm 1D} \simeq 2.4\pm0.4$ atoms \cite{comment1}.

In conclusion, we have used an integrated optical circuit with a photonic crystal waveguide to trap and interface atoms with guided photons.
Superradiance for atoms trapped along our APCW has been demonstrated and a peak single-atom emission rate into the APCW of $\Gamma_{\rm 1D}/\Gamma_0=1.1\pm0.1$ inferred. Our current uniform trap along the APCW is a promising platform to study optomechanical behavior induced by the interplay between sizable single-atom reflectivity and large optical forces (e.g., self organization~\cite{chang2013,grieser2013}).
By optimizing the power and detuning of an auxiliary guided mode field near the air band of the APCW, it should be possible to achieve stable atomic trapping and ground state cooling \cite{thompson2012,kaufman2012} at trap sites centered within the vacuum gap, thereby increasing $\Gamma_{\rm 1D}$ five-fold \cite{Hung13}. 
Opportunities for new physics in the APCW arise by fabricating devices with the atomic resonance inside the band gap to induce long-range atom-atom interactions~\cite{Shahmoon13, Douglas13, Tudela14}, thereby enabling investigations of novel quantum transport and many-body phenomena.

\textit{Acknowledgements} We gratefully acknowledge the contributions of D. J. Alton, D. E. Chang, K. S. Choi, J. D. Cohen, J. H. Lee, M. Lu, M. J. Martin, A. C. McClung, S. M. Meenehan, L. Peng, and R. Norte. Funding is provided by the IQIM, an NSF Physics Frontiers Center with support of the Moore Foundation, and by the DoD NSSEFF program (HJK), the AFOSR QuMPASS MURI, NSF PHY-1205729 (HJK) and the DARPA ORCHID program.
AG is supported by the Nakajima Foundation. SPY and JAM acknowledge support from the International Fulbright Science and Technology Award.

\pagebreak
\widetext
\begin{center}
\textbf{\large Supplemental Material: Superradiance for atoms trapped along a photonic crystal waveguide}
\end{center}

\setcounter{equation}{0}
\setcounter{figure}{0}
\setcounter{table}{0}
\setcounter{page}{1}
\makeatletter
\renewcommand{\theequation}{S\arabic{equation}}
\renewcommand{\thefigure}{S\arabic{figure}}
\renewcommand{\bibnumfmt}[1]{[S#1]}
\renewcommand{\citenumfont}[1]{S#1}

\section{Device characterization}
\label{sec:smdev}

\begin{figure}[b!]
\centering
\includegraphics[width=0.85\columnwidth]{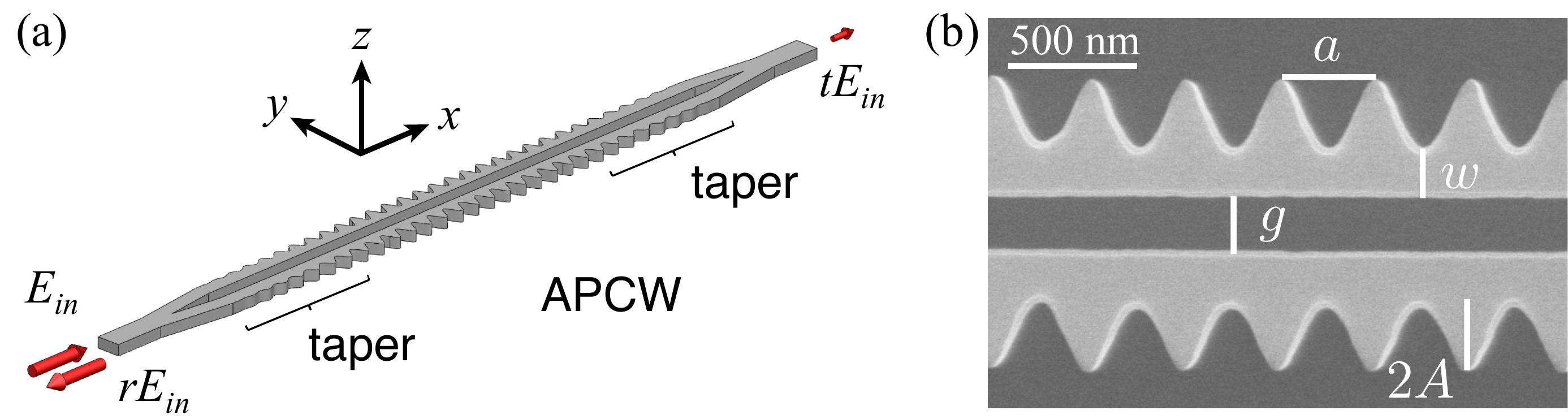}  
\caption{(a) Schematic of the APCW. An incident field $E_{\rm in}$ excites the TE-like fundamental mode, and the intensities fro the transmitted $tE_{\rm in}$ and reflected field $rE_{\rm in}$ are recorded for device characterization. (b) SEM image of APCW with lattice constant $a$, gap $g$, width $w$, and tooth amplitude $A$.}
\label{fig:APCW}%
\end{figure}

A schematic of the alligator photonic crystal waveguide (APCW) is illustrated in Fig. \ref{fig:APCW}(a). The waveguide is made from 200-nm thick stoichiometric SiN with refractive index $n=2.0$ \cite{Yu2014}. The dimensions of the nominal photonic crystals are the following: lattice constant $a = 371$ nm, gap $g = 238$ nm, width $w = 157$ nm, and tooth amplitude $A  = 131$ nm, as shown in Fig. \ref{fig:APCW}(b).  The nominal photonic crystal section consists of $N_{\rm cell}=150$ unit cells, terminated by 30 tapered cells on each side to provide `mode-matching' to and from double nanobeams sections.  
 
\begin{figure}
\centering
\includegraphics[width=0.8\columnwidth]{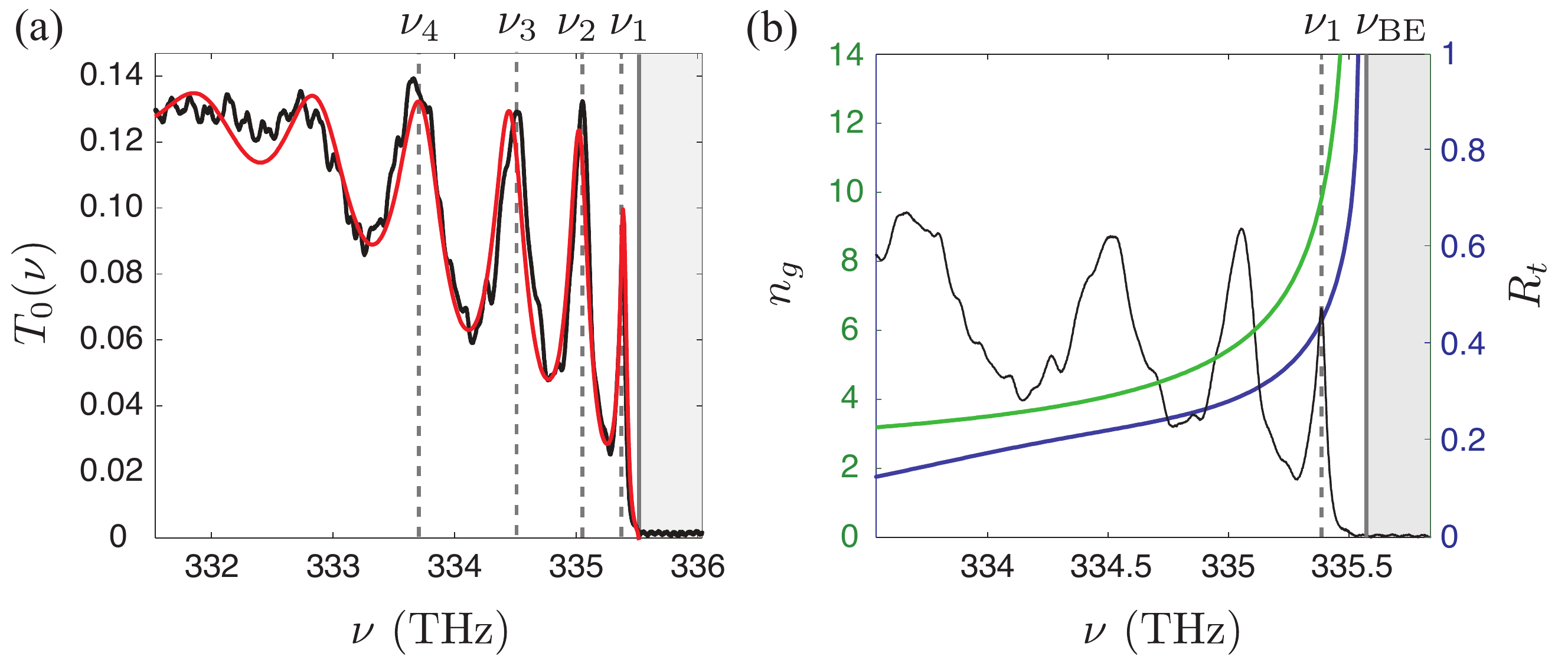}  
\caption{(a) Measured transmission spectrum $T_0(\nu)$ for the APCW (black) around the edge of the dielectric band and the model fit (red). The dashed lines mark the resonant frequencies $\nu_i$ from reflections in the taper sections and the solid line marks the band edge frequency $\nu_{\rm BE}$. (b) Estimated group index $n_g$ (green) and taper reflection $R_t$ (blue) from the fitted model. For the reference, the transmission spectrum $T_0(\nu)$ is overlaid. At the first resonance $\nu_1$ marked by the dashed line, the group index is $n_g\approx 11$, and the taper reflection is $R_t\approx 0.48$.}
\label{fig:smfig_ngandR}%
\end{figure}

The APCW is characterized by measuring the transmission spectrum $T_0(\nu)$ without atoms.
The resonant structure around frequencies $\nu_i$ displayed in Fig. \ref{fig:smfig_ngandR} arises from reflections in the tapered sections at the two ends of the APCW. The free spectral range $\Delta \nu_i = \nu_{i+1} -\nu_i$ between resonances decreases as the band edge frequency $\nu_{BE}$ is approached, which is a signature of an increasing group $n_g$ index near $\nu_{BE}$, with $n_{g} \propto1/\Delta \nu_i $ for an ideal structure.  

Our experiment in Ref. \cite{manuscript} is operated around the frequency $\nu_a$ of the D$_1$: 6$S_{1/2},~F=3\rightarrow 6P_{\rm 1/2},~ F'=4$ transition in atomic Cs, with $\nu_A$ aligned near $\nu_{BE}$ by absolute control of the fabrication process at a level of $10^{-3}$. Fine tuning for $\nu_A = \nu_1$ is achieved by way of a guided-mode (GM) heating beam with a wavelength of 850 nm and optimum power, typically $P \geq 100\mu$W. In addition, we turn on a strong GM heating beam for 100 ms at the end of each experimental cycle in order to keep the device clean by desorbing Cs from the APCW.  

In order to estimate the group index $n_g$,  single-taper reflectivity $R_t$, and intensity loss $e^{-2\zeta}$, we use a model based on the transfer matrix formalism for a periodic system to fit the transmission spectrum~\cite{Hood2015}, which we now briefly describe. 
The dispersion relation for the wavevector $k(\nu)$ near the band edge is approximated by the fitting function~\cite{Hood2015},
\begin{equation}
k(\nu) = k_0 \left( 1-\sqrt{ \frac{\left(\nu_0 - \nu(1+i\kappa) \right)^2-\Delta_g^2}{ \nu_\text{F}^2-\Delta_g^2 }  }       \right),
\label{eq:k}
\end{equation}
where the wavevector at the band edge is $k_0=\pi/a$. 
Here, fitting parameters are the frequency at the center of the band gap $\nu_0$, the size of the band gap $2\Delta_g$, the asymptotic group velocity far from the band edge $ 2 \pi \nu_\text{F}/k_0 $, and the loss parameter $\kappa$. 
The loss parameter $\kappa$ comes from using perturbation theory to add a small imaginary component to the dielectric constant of the material, resulting in an imaginary propagation constant that is approximately given by
\begin{equation}
\text{Im}[k(\nu)] \approx  \frac{ 2 \pi \nu}{v_\text{g}} \kappa .
\label{eq:}
\end{equation}
It provides a convenient way to model losses that scale with inverse group velocity.

Next we consider the weak cavity formed by the taper reflections $R_t$. The single-pass phase accumulation $\phi$ and single-pass power transmission $e^{-2\zeta}$ through the cavity are written by 
\begin{equation}
\phi = N_{\rm cell}  a\, \text{Re}[k]   \quad \text{and} \quad  \zeta = N_{\rm cell} a \, \text{Im}[k],
\label{eq:}
\end{equation}
where $N_{\rm cell}$ is the number of unit cells of the APCW and $a$ is the lattice constant. 
Then, the transmission through a symmetric cavity with mirrors $R_t$
is given by
\begin{equation}
T_\text{cavity} = \frac{1}{1+\mathcal{L} + F \sin^2[\phi]},
\label{eq:Tcavity}
\end{equation}
where the coefficient $F$ 
and loss coefficient $\mathcal{L}$ are given by 
\begin{equation}
\mathcal{L} = \frac{\left( 1-R_t \, e^{-2 \zeta} \right)^2 }{e^{-2 \zeta} \, (1-R_t)^2} -1 \quad \text{ and } \quad  F = \frac{4 R_t}{(1-R_t)^2 }.
\label{eq:}
\end{equation}

In order to fit this model to the measured transmission spectrum, first we use the dispersion model (Eq. (\ref{eq:k}) with $\kappa=0$) to fit the positions of the cavity resonances to. Second, we fit Eq. \eqref{eq:Tcavity} with no loss ($\mathcal{L}=0$) to the transmitted spectrum by using the fitted dispersion model to find $\phi$ and by using a fitting function for $F$ \cite{Hood2015}, namely
\begin{equation}
(F)^{-1/2} = A_1 (\mathrm{d}\nu/\Delta_g) + A_2 (\mathrm{d}\nu/\Delta_g)^2 + A_3 (\mathrm{d}\nu/\Delta_g)^3  ,
\label{eq:}
\end{equation}
where $d\nu$ is the distance in frequency from the band edge. 
Finally, we find  the loss parameter $\zeta$ that makes the on-resonant peak heights of the model best match our measurement. 

Figure \ref{fig:smfig_ngandR} (a) shows the measured transmission spectrum (black curve), overlaid with the model fit (red curve). The fitted parameters for the dispersion model are $2 \Delta_g$ = 14.44 THz, $\nu_\text{F}/\nu_0$ = 0.60, and $\nu_0$ = 342.8 THz. The fitted parameters for $F$ are $A_1$ = 9, $A_2$ =-48, and $A_3$ = 128. The fitted loss parameter is $\kappa = 1.5\times 10^{-5} $.  At the first resonance $\nu_1$, the model linewidth is 55 GHz, in reasonable agreement with the measured linewidth of 66 GHz.
The fitted dispersion relation is used to estimate the group index, and the fitted cavity model is used to estimate $R_t$ and the single pass transmission $e^{-2 \zeta}$, as shown in Fig. \ref{fig:smfig_ngandR} (b).  At the first resonance, the group index is $n_g\approx 11$, the taper reflection is $R_t\approx 0.48$, resulting in a peak intensity enhancement $\mathcal{E}_I = \frac{1+\sqrt{R_t}}{1-\sqrt{R_t}}\approx 5.5$ \cite{Hood2015}, and the single-pass transmission is $e^{-2\zeta} \approx 0.89 $.  Since the propagation loss in the APCW is reasonably small, we ignore the loss in our analysis in the following sections.

\section{Finite different time domain calculations for collective coupling rates}
 \label{sec:fdtd}
 \begin{figure}
\centering
\includegraphics[width=.45\columnwidth]{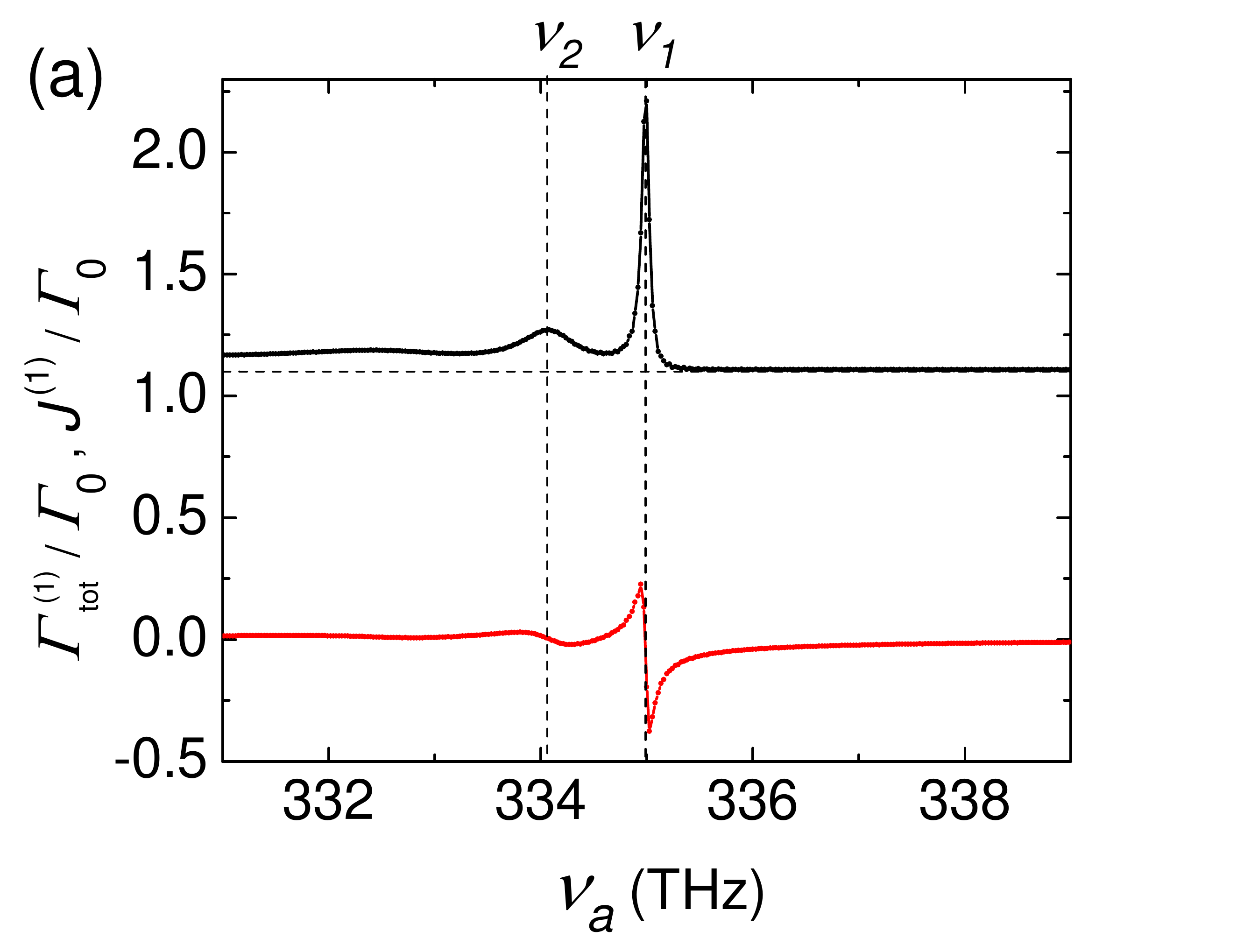}
\includegraphics[width=.45\columnwidth]{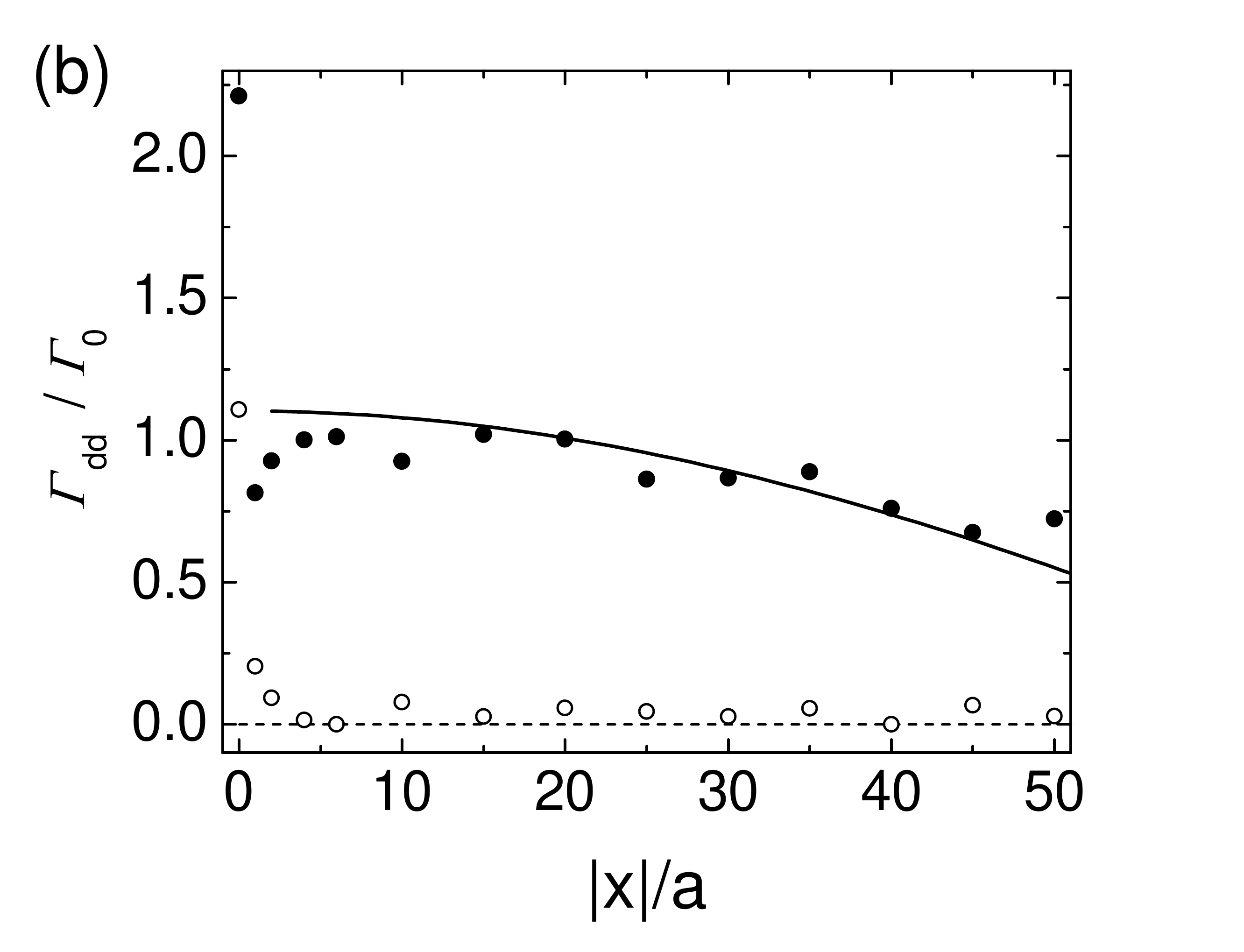}
\caption{(a) Single-atom decay rate $\Gamma^{(1)}_\mathrm{tot}$ (black circles) and excited state level shift $J^{(1)}$ (red circles) at $\mathbf{r}_a = (0,0,z_1)$~nm. Vertical dashed lines mark the frequencies of the first two guided mode resonances, $\nu_1$ and $\nu_2$, near the band gap (frequency range $\nu_a \gtrsim \nu_\mathrm{BE} = 335.5~$THz where $\Gamma_\mathrm{tot}$ appears constant) that are supported by the finite length of the APCW. Horizontal dashed line indicates $\Gamma'/\Gamma_0 = 1.1$, estimated from the constant $\Gamma^{(1)}_\mathrm{tot}$ in the band gap region.  (b) Dissipative coupling rate $\Gamma_\mathrm{dd} (x) \equiv |\Gamma(\mathbf{r}_a,\mathbf{r}_a+x\hat{\mathbf{x}})|$ between two trapped atoms separated by $x$, with their resonant frequencies at either $\nu_a =\nu_1$ (solid circles) or $\nu_a = 336~$THz $>\nu_{\rm BE}$  inside the band gap (open circles), respectively. Solid line is an analytical calculation considering actual finite size of the APCW (Fig. \ref{fig:APCW}).}
\label{smfig_coupling}
\end{figure}

Due to strong coupling to the TE-like GM in the APCW, trapped atoms experience both enhanced atomic decay rates as well as collective Lamb shifts. To estimate the size of these effects, we perform FDTD calculations and Fourier analysis as described in Ref.~\cite{Hung13s} to obtain the two-point Green's tensor $\mathbf{G}(\mathbf{r}_1,\mathbf{r}_2,\omega)$ for the APCW shown in Fig. \ref{fig:APCW}. We then evaluate dissipative and coherent  coupling rates, respectively, as \cite{SMAgarwal75, SMBuhmann04,SMBuhmann08}
\begin{eqnarray}
\Gamma (\mathbf{r}_1,\mathbf{r}_2) &=& \frac{2\mu_0 \omega_a^2}{ \hbar}  \mathbf{d} \cdot \mathrm{Im}[\mathbf{G}(\mathbf{r}_1,\mathbf{r}_2,\omega_a)] \cdot \mathbf{d} \\ \label{decay}
J (\mathbf{r}_1,\mathbf{r}_2)&=& -\frac{\mu_0 \omega_a^2}{\hbar} \mathbf{d}\cdot \mathrm{Re} [\mathbf{G}_{sc}(\mathbf{r}_1,\mathbf{r}_2,\omega_a)] \cdot \mathbf{d},
\end{eqnarray} 
where $\mathbf{d}$ is the transition dipole moment, $\omega_a$ the transition frequency, $\mu_0$ the vacuum permeability, and $\hbar$ Planck's constant divided by $2\pi$. Here, $\mathbf{G}_{sc} = \mathbf{G} - \mathbf{G}_0$ is the scattering Green's tensor, in which the vacuum contribution $\mathbf{G}_0$ is subtracted from the total Green's tensor $\mathbf{G}$; $\mathrm{Im}[.]$ and $\mathrm{Re}[.]$ represent imaginary and real parts, respectively. The coupling rate $\tilde{\Gamma}= \Gamma/2 + i J$ controls collective excitation dynamics of trapped atoms along the APCW.

We obtain single-atom rates by setting $\mathbf{r}_1=\mathbf{r}_2=\mathbf{r}_a$ at the location of a trapped atom, and evaluate the single-atom total decay rate $\Gamma_{\mathrm{tot}}^{(1)} (\nu_a)= \Gamma(\mathbf{r}_a,\mathbf{r}_a,\nu_a)$ and excited state level shift $J^{(1)} (\nu_a)= J(\mathbf{r}_a,\mathbf{r}_a,\nu_a)$. Figure~\ref{smfig_coupling}(a) shows the calculation for $\mathbf{r}_a = (0,0,z_1)$~nm at the center of the trap shown in Fig. 1(b) of \cite{manuscript}. Here the total decay rate $\Gamma^{(1)}_{\mathrm{tot}} = \Gamma_\mathrm{1D} + \Gamma'$ (black curve) includes the contribution from the GM of interest ($\Gamma_\mathrm{1D}$), which strongly depends on the atomic resonant frequency $\nu_a=\omega_a/2\pi$ and position $\mathrm{r}_a$, as well as the coupling rate to all other modes ($\Gamma'$). $\Gamma'$ can be estimated from $\Gamma^{(1)}_\mathrm{tot}(\nu_a)$ inside the band gap ( $\nu_a \gtrsim \nu_\mathrm{BE}$); $\Gamma'/\Gamma_0  \approx 1.1$ remains constant over a broad frequency range. Coupling rate to the TE-like GM, $\Gamma_\mathrm{1D} =\Gamma^{(1)}_{\mathrm{tot}} - \Gamma'  $, can be obtained from this analysis with $\Gamma_{\rm 1D}/\Gamma_0=1.2$.

In Fig.~\ref{smfig_coupling} (a), we calculate a small excited state level shift $|J^{(1)}|/\Gamma_0 < 0.4$ over a frequency range around $\nu_1 = 335~$THz. For our experimental configuration, with $\nu_a \approx \nu_1$, we find $|J^{(1)}(\nu_a)|/\Gamma_0 \sim 0 $. This also suggests that the collective level shift for two trapped atoms, $|H_{\rm dd}(x)| \equiv |J(\mathbf{r}_a,\mathbf{r}_a+x\hat{\mathbf{x}})| \ll(\Gamma_0, \Gamma_\mathrm{1D})$, is negligible, where $x$ is the atomic separation. Indeed, we do not see clear evidence of $N$-dependent level shifts in the steady-state transmission spectra shown in Fig.~4(a) of \cite{manuscript}.

Figure~\ref{smfig_coupling} (b) shows $\Gamma_\mathrm{dd} (x) \equiv |\Gamma(\mathbf{r}_a,\mathbf{r}_a+x\hat{\mathbf{x}})|$ for two trapped atoms located at the center of unit cells ($x/a \in \mathbb{Z}$) and with resonant frequencies at $\nu_a = \nu_1~$ or $\nu_a > \nu_\mathrm{BE}$ inside the band gap, where $\Gamma^{(1)}_\mathrm{tot} - \Gamma' \sim 0$. For $|x| > a$, $\Gamma_\mathrm{dd} (x)$ can be used to estimate the dissipative coupling rate between two atoms. When $\nu_a  =\nu_1$ and $|x|/a>2$, $\Gamma_\mathrm{dd} (x)$ slowly drops from $\Gamma_\mathrm{dd}(0)-\Gamma' =1.2~\Gamma_0$ to smaller values as $|x|$ becomes comparable to the size of the APCW (black circles). This is caused by interference with reflections from the tapering regions surrounding the APCW. 
Solid line in Fig.~\ref{smfig_coupling}(b) shows an analytical calculation $\Gamma_\mathrm{dd}(x) =(\Gamma_\mathrm{dd}(0)-\Gamma') \cos (\pi |x|/N_{\rm eff}a)$ that compares to the numerical result, where the fitted effective number of cells $N_{\rm eff}=162\pm9$ is larger than $N_{\rm cell}$ due to the leakage of the fields into the taper regions.
Small variations between the analytical and numerical calculations are due to residual coupling via other channels. On the other hand, when $\nu_a>\nu_\mathrm{BE}$  inside the band gap, $\Gamma_\mathrm{dd} (x)$ quickly drops below $ 0.1\Gamma_0$ at $|x|/a>2$. This is expected because, inside the band gap, atoms can only cooperatively decay via photonic channels that contribute to $\Gamma'$, which are either weakly-coupled or are lost quickly into freepace within distances $|x|<2a$.

\section{Lifetime of trapped atoms along the APCW}
\begin{figure}
\centering
\includegraphics[width=.8\columnwidth]{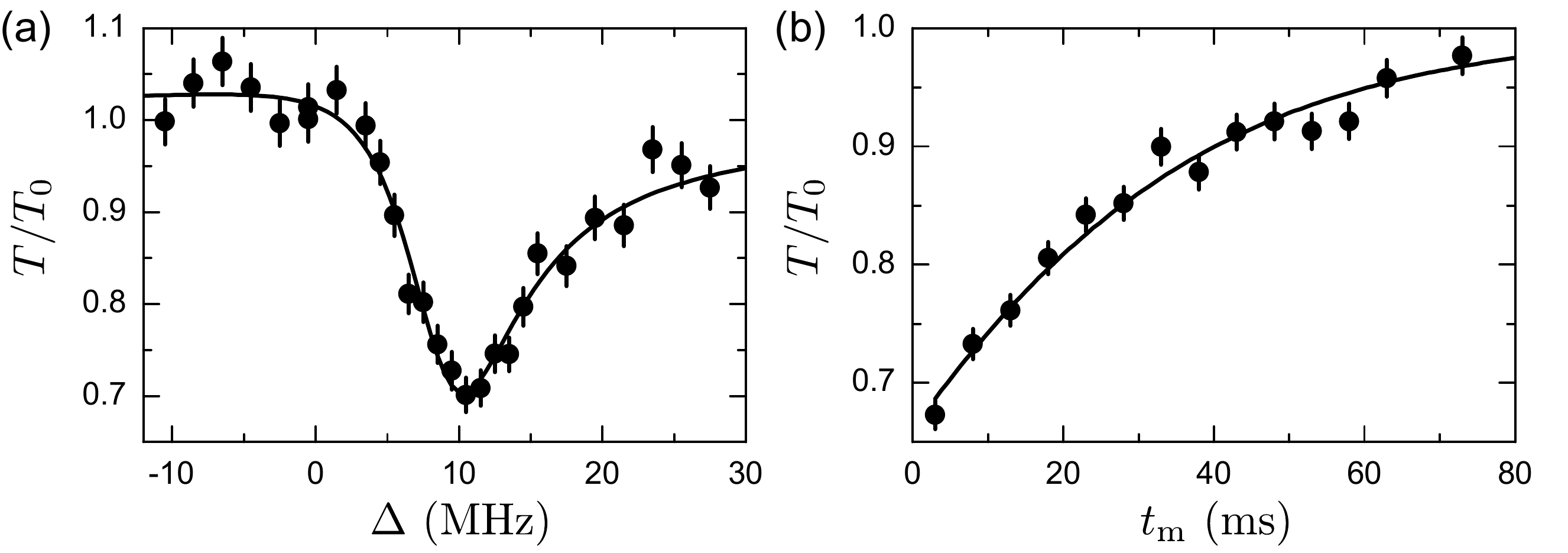}
\caption{(a) Normlaized transmission spectrum. Black curve shows the fit to Eq. (\ref{eq:sseq}) at $t_{\rm m}=2.5$ ms with $\theta=-0.6\pm0.1$, $C_0=0.24\pm0.1$, $\Gamma=8.2\pm0.6$ MHz and $\Delta_0=9.3\pm0.3$ MHz. (b) Normalized transmission $T/T_0$ as a function of the holding time, measured by the on resonant guided-mode probe with $\Delta=10.5$ MHz. By fitting the measured data to Eq. (\ref{eq:sseq}) with fitted parameters extracted in Fig. \ref{smfig2} (a), we obtain the lifetime of $\tau_{\rm GM}=28\pm2$ ms (black curve).}
\label{smfig2}
\end{figure}

To characterize the lifetime of trapped atoms near the APCW, we measure the normalized transmission $T/T_0$ as a function of the measurement time  $t_{\rm m}$, as shown in Fig. 2~\cite{manuscript} and replotted in Fig. \ref{smfig2}. During the lifetime measurement, the frequency $\nu_{a}$ of the D$_1$ transition for the probe field $E_{\rm in}$ is located between the first and second taper resonances, which leads to the dispersive spectrum shown in Fig \ref{smfig2} (a). In order to estimate the lifetime of the trap with off resonant cavity, we employ the steady state equation~\cite{Thompson1998}, 
\begin{eqnarray}
T/T_0=(1+\theta^2)/\left[\left(1+\frac{2C(t_{\rm m})}{1+\delta_{\rm m}^2}\right)^2+\left(\theta-\frac{2C(t_{\rm m})\delta_{\rm m}}{1+\delta_{\rm m}^2}\right)^2\right],
\label{eq:sseq}
\end{eqnarray}
where the normalized detuning from the light shifted resonance $\Delta_0$ is $\delta_{\rm m}=\frac{\Delta-\Delta_0}{\Gamma}$, the cooperativity parameter is  $C(t_{\rm m})=C_0\exp(-t_{\rm m}/\tau_{\rm GM})$ with peak cooperatively $C_0$, lifetime $\tau_{\rm GM}$, and normalized detuning from taper resonance $\theta$. First, we fit the measured spectrum at $t_{\rm m}=2.5$ ms to Eq. (\ref{eq:sseq}) and obtain the fitted parameters, $\theta=-0.6\pm0.1$, $C_0=0.24\pm0.1$, $\Gamma=8.2\pm0.6$ MHz and $\Delta_0=9.3\pm0.3$ MHz as shown in black curve in Fig. \ref{smfig2} (a). 
Then, to estimate the lifetime $\tau_{\rm GM}$, the measured data for $\Delta=10.5$ MHz in Fig. \ref{smfig2} (b) are fitted to Eq. (\ref{eq:sseq}) with fitted parameters from Fig. \ref{smfig2} (a). We obtain the lifetime of $\tau_{\rm GM}=28\pm2$ ms shown in black curve in Fig. \ref{smfig2} (b).

\section{Model for superradiance of trapped atoms}
\label{sec:model}
Our model of superradiance of trapped atoms is obtained by including transfer matrices for atoms in the device model described
in Section \ref{sec:smdev}~\cite{Deutsch1995s, Chang2007bs, Chang12s, Hood2015}. Since the first resonance of the taper reflections is aligned to the D$_1$ transition for the probe $E_{\rm in}$, the wavevector  of the probe mode is $k=\left(1-\frac{1}{N_{\rm eff}}\right)\frac{\pi}{a}$ with number of cells $N_{\rm eff}$, and the probe field inside the unit cell forms a nearly perfect standing wave due to the Bloch-periodic function.
In addition, atoms are trapped near the central region of the APCW along the $x$ axis ($\Delta x=\pm10~\mu$m). Thus, we ignore the dephasing between atoms and envelope from the taper reflections due to the small mismatch of the wavevector $\Delta k=\frac{1}{N_{\rm eff}}\frac{\pi}{a}$ relative to $k_0=\frac{\pi}{a}$ at the band edge. In the following, we set the wavevector $k=k_0=\frac{\pi}{a}$ and will discuss effects due to the $\Delta k$ mismatch later.

\begin{figure}[t!]
\centering
\includegraphics[width=.8\columnwidth]{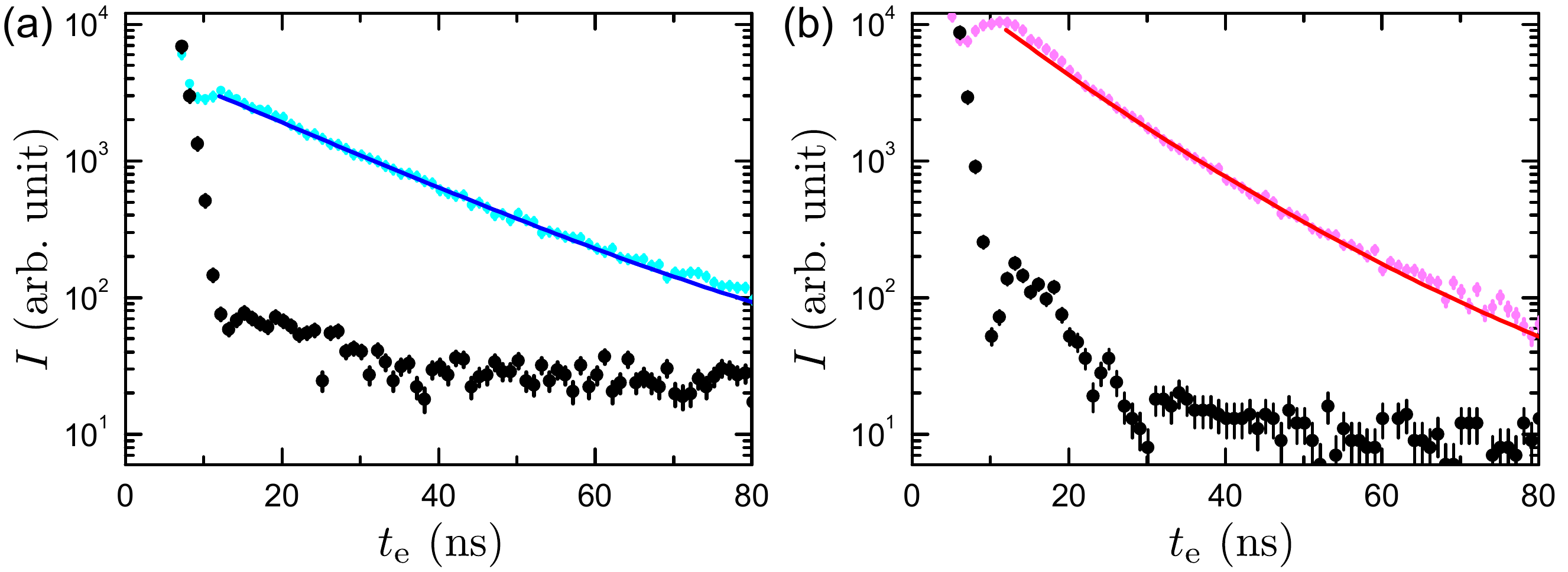}
\caption{(a) Temporal profiles of atomic emission into the GM at $t_{\rm m}=63$ ms with and without atoms, shown in cyan and black circles, respectively. The blue curve shows Eq. (\ref{eq:argcell}) fitted for a single atom with $\Gamma_{\rm 1D}/\Gamma'=1.0\pm0.1$. (b) Temporal profiles of atomic emission into the GM at $t_{\rm m}=3$ ms with and without atoms, shown in pink and black circles, respectively. The red curve shows Eq. (\ref{eq:argcellpos}) fitted to yield with $\bar{N}=2.6\pm0.3$. The background level of (a) is higher than (b) due to the drift of the intensity modulator during the 5 times longer data accumulation time.}
\label{smfig_fit}
\end{figure}

The reflection of $N$ atoms randomly distributed at the location $x_i$ with the coupling rate $\Gamma_{\rm 1D}\cos^2(kx_i)$ is given by, 
\begin{equation}
r_{N}(\delta)=\frac{i\xi_{N}}{1-i\xi_{N}}~~~~~~{\rm where}~~~~~~\xi_{N}=-\frac{\xi_0}{i+\delta}\sum_i\cos^2\left(kx_i\right).
\label{eq:nr}
\end{equation}
where the single-atom fractional coupling rate is $\xi_0=\Gamma_{\rm 1D}/\Gamma'$ and normalized detuning is $\delta=2\Delta/\Gamma'$ with $\Gamma'/\Gamma_0\approx1.1$ from the numerical simulation in Section \ref{sec:fdtd}~\cite{Hung13s}.
The temporal profile of superradiance from $N$ atoms is obtained by Fourier transforming $r_N(\delta)$ to yield $r_N(t)$, and taking the convolution of $r_N(t)$ with a gaussian pulse of the half width $\sigma\sim5$ ns for the excitation pulse $E_{\rm in}(t)$.
Furthermore, the temporal profile at $t>2\sigma$ can be approximated by
\begin{equation}
I_{r^{\rm conv}_N}(t)=|r_N^{\rm conv}(t)|^2\propto\Big(\Gamma_{\rm 1D}\sum_{i}\cos^2\left(kx_i\right)\Big)^2\cdot\exp\left[-\left(\sum_i\Gamma_{\rm 1D}\cos^2\left(kx_i\right)+\Gamma'\right)t\right].
\label{eq:sp}
\end{equation}
Considering the assumed random locations of atoms with uniform probability density in the unit cells along $x$, the spatially averaged temporal profile is obtained by integrating Eq. (\ref{eq:sp}) along $x$, yielding
\begin{eqnarray}
\mathcal{I}_N(t)&=&\gamma^2 e^{-(N\gamma+\Gamma')t}\cdot{I_{0}}\left(\gamma t\right)^{N-2}\cdot\left[\frac{N(N+1)}{4}{I_0}\left(\gamma t\right)^2-\left(\frac{N}{4 \gamma t}+\frac{N^2}{2}\right){I_{0}}\left(\gamma t\right){I_{1}}\left(\gamma t\right)+\frac{N(N-1)}{4}{I_{1}}\left(\gamma t\right)^2\right],
\label{eq:argcell}
\end{eqnarray}
where $I_k(z)$ is a modified Bessel function of the first kind and $\gamma=\Gamma_{\rm 1D}/2$. 
In addition, the number of trapped atoms along the APCW is drawn from a Poisson distribution $p(\bar{N},N)$ with mean number of atoms $\bar{N}$. The total decay curve then becomes
\begin{eqnarray}
\mathcal{I}_{\rm tot}(t)=c_0\sum_N p(\bar{N},N)\cdot\mathcal{I}_{N}(t)+ \mathcal{I}_{\rm BG},
\label{eq:argcellpos}
\end{eqnarray}
where $c_0$ is a constant. Here, the background intensity $\mathcal{I}_{\rm BG}$ is measured separately without atoms and is given by the black circles in Fig. \ref{smfig_fit}.
As shown in Fig. 3~\cite{manuscript}, the total decay rate asymptotes to $\bar{\Gamma}^{(1)}_{\rm tot}/\Gamma_0=2.0\pm0.1$ at the longer hold times, which suggests that atomic decay at $t_{\rm m}=63$ ms mostly originates from a single atom for mean atom number $\bar{N}\ll1$.
Thus, we fit the decay curve at $t_{\rm m}=63$ ms to Eq. (\ref{eq:argcell}) with $N=1$, and obtain $\xi_0=\Gamma_{\rm 1D}/\Gamma'=1.0\pm0.1$ shown in Fig. \ref{smfig_fit} (a).
Then, by using the fitted $\xi_0$, the shortest hold time data ($t_{\rm m}=3$ ms) is reasonably well fitted to Eq. (\ref{eq:argcellpos}) with $\bar{N}=2.6\pm0.3$, as shown in Fig. \ref{smfig_fit} (b). 

We also numerically estimate the contribution of the envelope from the taper reflections and dephasing between atoms along the APCW. 
We employ the transfer matrix model with $k=k_0-\Delta k$, which includes the coupling rate $\Gamma_{\rm 1D}(x)\simeq\Gamma^{(0)}_{\rm 1D}\cos^2(k_0x)\cos^2(\Delta kx)$ along the $x$-axis of the APCW and propagation phase $\Delta k \cdot\delta x$ between atoms separated by $\delta x$. Here, $\Gamma^{(0)}_{\rm 1D}$ denotes the peak coupling rate of both the unit cell and the envelope from taper reflections. The position of the atoms is generated from a normal distribution with $\sigma_{x}=10~\mu$m at the temperate of $50\mu$K. Then, we numerically generate the decay curve and extract the total decay rate, suggesting that $\Gamma_{\rm 1D}$ extracted from Eq. (\ref{eq:argcell}) with $N=1$ is underestimated by $\sim10$\% and $\bar{N}$ from the fits to Eq. (\ref{eq:argcellpos}) by $\sim15$\%. Note that we do not incorporate these corrections in our estimation of $\Gamma_{\rm 1D}$ and $\bar{N}$ in Ref.~\cite{manuscript}, since the temperature of the atoms trapped along the APCW could be different from the measured temperature in free space. Indeed, the calculated trap potential combined with Casimir-Polder potential suggests that atoms trapped along the APCW could be much colder ($\lesssim20~\mu$K) than the measured temperature for the free-space FORT ($\sim 50~\mu$K) due to the smaller trap depth near the APCW, leading to a smaller  correction of $\Gamma_{\rm 1D}$ and $\bar{N}$ due to tighter localization of the atoms around the center of the APCW. The distribution of atoms along the APCW is being investigated in more detail.

\begin{figure}[h!]
\centering
\includegraphics[width=.45\columnwidth]{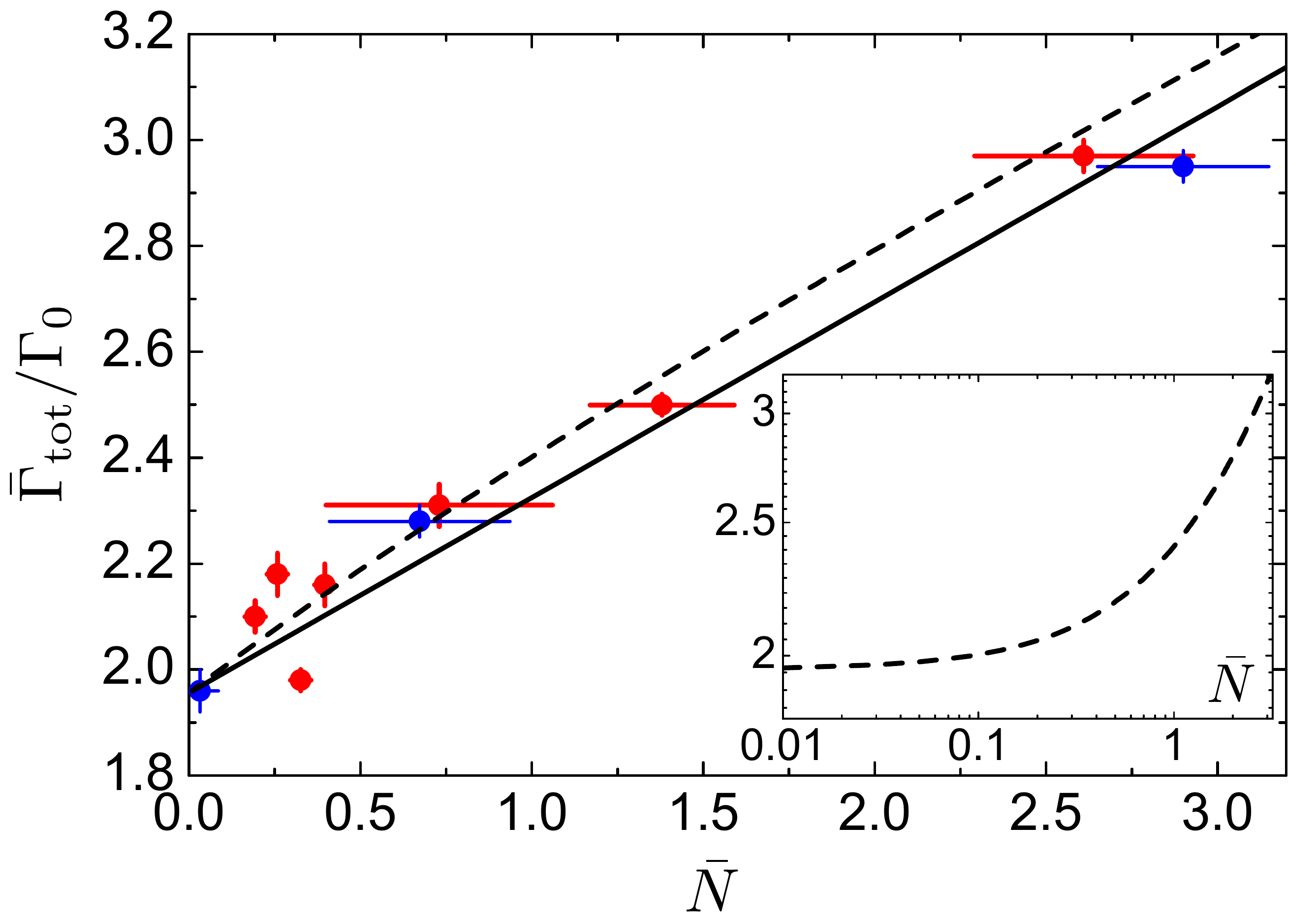}
\caption{Fitted total decay rate $\bar{\Gamma}_{\rm tot}$ normalized by $\Gamma_0$ as a function of the mean number of trapped atoms $\bar{N}$.
The dashed curve shows the calculated $\bar{\Gamma}_{\rm tot}$ from the model, overlaid with results measured for various hold times (red circles) and for loading times (blue circles). A linear fit to the combined data (solid black line) gives $\bar{\Gamma}_{\rm SR}=\eta\cdot\bar{N}\cdot\Gamma_{\rm 1D}$ with $\eta=0.34\pm0.06$. The inset shows a $\log$-$\log$ plot of the curve generated from the model.}
\label{smfig_GtotN}
\end{figure}

To support our assumption of $\bar{N}$-dependent superradiance, $\bar{\Gamma}_{\rm tot}=\bar{\Gamma}_{\rm SR}+\bar{\Gamma}_{\rm 1D}$ with $\bar{\Gamma}_{\rm SR}=\eta\cdot\bar{N}\cdot\Gamma_{\rm 1D}$, we generate the decay curve from Eq. (\ref{eq:argcellpos}) with $\Gamma_{\rm 1D}/\Gamma_0=1.0$ and various $\bar{N}$, and extract $\bar{\Gamma}_{\rm tot}$ by fitting to an exponential.  The dashed curve in Fig. \ref{smfig_GtotN} shows the calculated $\bar{\Gamma}_{\rm tot}$, overlaid with measured hold time (red circles) and loading time (blue circles) dependence and the linear fit (solid black line).
Although the dashed curve generated from the model deviates from the linear dependence at $\bar{N}<1$, the linear fit captures the $\bar{N}$ dependence reasonably well. As clearly seen in the inset of Fig. \ref{smfig_GtotN}, the nonlinear dependence on $\bar{N}$ at $\bar{N}\ll1$ is due to the ``conditional" character of decay rate measurements, meaning that the decay curve consists mostly of fluorescence from a single atom, despite $\bar{N}\ll1$.
Due to the negligible background counts in our measurements, single detection events at $\bar{N}\ll1$ herald the presence of single atoms.

The linear fit to the combined data sets gives $\bar{\Gamma}_{\rm SR}=\eta\cdot\bar{N}\cdot\Gamma_{\rm 1D}$ with $\eta=0.34\pm0.06$, consistent with the model for $\bar{N} \gtrsim 0.7$.
A qualitative understanding of this value of $\eta$ is the following. 
Due to the random distribution of atoms along the APCW, the intensity of atomic emission into the GM is spatially modulated by $\cos^4(kx)$ as shown in Eq. (\ref{eq:sp}), meaning that both GM excitation of atoms and emission into the GM are proportional to $\cos^2(kx)$, resulting in $\cos^4(kx)$ dependence.
With the spatial averaging along the $x$-axis of the APCW, the superradiant decay rate is then reduced by a factor of roughly $\eta\sim3/8$ (i.e., the average of $\cos^4(kx)$ over a unit cell).



\begin{thebibliography}{99}
\bibitem{comment1} $\bar{N}_{\rm m}$ is defined from $\bar{\Gamma}_{\rm m}(\rho)=\bar{N}_{\rm m}\bar{\Gamma}_{\rm 1D}+\Gamma'$ for an approximate estimate of the number of atoms. Empirically we find $\bar{N}_{\rm m}\sim\bar{N}$.
\end{thebibliography}
\end{document}